\documentclass{aa}
\usepackage{psfig}
\def\solar {\ifmmode_{\mathord\odot} \else $_{\mathord\odot}$\fi}
\def\Msol {\ifmmode {\,{\rm M}\solar} \else $\,{\rm M}$\solar\fi} 
\def\Rsol {\ifmmode {\,{\rm R}\solar} \else $\,{\rm R}$\solar\fi} 

\newcommand{\kms}{\,km\,s$^{-1}$\,} 
\begin{document}
\title{
New neighbours. III. 21 new companions to nearby dwarfs, discovered with 
  adaptive optics
  \thanks{
Based on observations made at Canada-France-Hawaii Telescope, operated by
the National Research Council of Canada, the Centre National de la
Recherche Scientifique de France and the University of Hawaii. Some of
the data presented herein were obtained at the W.M. Keck Observatory, which
is operated as a scientific partnership among the California Institute of
Technology, the University of California and the National Aeronautics and
Space Administration. The Observatory was made possible by the generous
financial support of the W.M. Keck Foundation.}} 
%
\author{J.-L.~Beuzit \inst{1,2}, 
        D.~S\'egransan \inst{1},
        T.~Forveille \inst{2,1}, 
        S.~Udry \inst{4}, 
        X.~Delfosse \inst{1,3}, 
        M.~Mayor \inst{4}, 
        C.~Perrier \inst{1}, 
        M.-C.~Hainaut \inst{2,5},
        C.~Roddier \inst{6},
        F.~Roddier \inst{6},
        E.L.~Mart\'{\i}n \inst{6}
       }
\offprints{Thierry.Forveille, e-mail: forveille@cfht.hawaii.edu}
\authorrunning{Beuzit et al.}
\institute{   Observatoire de Grenoble,
              Universit\'e J. Fourier, BP53,
              F-38041 Grenoble,
              France
\and
              Canada-France-Hawaii Telescope Corporation, 
              65-1238 Mamalahoa Highway,
              Kamuela, HI 96743, Hawaii
              U.S.A.
\and
              Instituto de Astrofisica de Canarias
              E-38200 La Laguna, Tenerife, 
              Canary Islands, Spain
\and
              Observatoire de Gen\`eve,
              CH-1290 Sauverny,
              Switzerland
\and
              now at, Gemini Observatory, Southern Operations Center,
              c/o AURA Inc., Casilla 603, La Serena,
              Chile  
\and
              Institute for Astronomy, 
              University of Hawaii, 
              2680 Woodlawn Drive, Honolulu,
              HI 96822, Hawaii, U.S.A.
             }
\date{Received ; accepted }
\markboth{Beuzit et al. }{22 new nearby dwarfs} 
\abstract{
We present some results of a CFHT adaptive optics search
for companions to nearby dwarfs. We identify 21 new components
in solar neighbourhood systems, of which 13 were found while surveying
a volume-limited sample of M dwarfs within 12~pc. 
We are obtaining complete observations for this
subsample, to derive unbiased multiplicity statistics for the 
very-low-mass disk population. Additionally, we resolve for the first time 
6 known spectroscopic or astrometric binaries, for a total of 27 newly 
resolved companions. A fair fraction of the new binaries has favourable
parameters for accurate mass determinations. \\
The newly resolved companion of Gl~120.1C had an apparent 
spectroscopic minimum mass in the brown-dwarf range (Duquennoy \& Mayor
\cite{duquennoy91}), and it contributed to the statistical evidence 
that a few percent of solar type stars might have close-in brown-dwarf 
companions.
We find that Gl~120.1C actually is an unrecognised double-lined
spectroscopic pair. Its radial-velocity amplitude had therefore been strongly
underestimated by Duquennoy \& Mayor (\cite{duquennoy91}), and it 
does not truly belong to their sample of single-lined systems with
minimum spectroscopic mass below the substellar limit. \\
We also present the first direct detection of Gl~494B, an astrometric 
brown-dwarf candidate. Its luminosity does straddle
the substellar limit, and it is a brown dwarf if its age is less than
$\sim$300~Myr. A few more years of observations will ascertain its mass
and status from first principles.
}
\maketitle

\section{Introduction}

As discussed in more detail in the first paper of this series (Delfosse et al. 
\cite{delfosse99b}), stellar multiplicity is a key input for a number
of important astrophysical issues. The joint distributions of system masses, 
mass ratios, semi-major axes, and eccentricities represent powerful 
diagnostics of the formation and early dynamical evolution of stellar 
systems (e.g. Duch{\^e}ne \cite{duchene99}; Patience et al. \cite{patience98};
Bonnell et al. \cite{bonnell98}; Kroupa \cite{kroupa00}). For most stellar
classes, unresolved companions also represent the main uncertainty when 
deriving the mass or luminosity function (e.g. Kroupa 
\cite{kroupa00}).  

The multiplicity statistics are now fairly well determined for the G
and K dwarfs (Duquennoy \& Mayor \cite{duquennoy91}; Halbwachs, Mayor and 
Udry \cite{halbwachs98}), but considerably more uncertain for higher and 
lower-mass stars. For M dwarfs the samples are either
very small (34 M dwarfs within 5.2~pc; Henry \& McCarthy \cite{henry90}; 
Leinert et al. \cite{leinert97}), or they have significant and uncertain
completeness corrections (Reid \& Gizis, \cite{reid97}; Fisher \& Marcy
\cite{fisher92}). The M-dwarf binary fractions derived from these data 
range from 26\% (Leinert et al. \cite{leinert97}) through 35\% (Reid \& Gizis, 
\cite{reid97}) to 42\% (Fisher \& Marcy \cite{fisher92}). This points 
towards a smaller fraction of multiple stars than the 57\% found 
amongst G dwarfs (Duquennoy \& Mayor, \cite{duquennoy91}), but still
with relatively modest significance ($\sim$3~$sigma$ for the 5.2~pc sample).
The parameter distributions, are determined only from the subsamples of
identified binaries, and thus even more uncertain. As just one consequence, 
contrasting assumptions on
stellar multiplicity (Kroupa \cite{kroupa95}; Reid \& Gizis \cite{reid97})
can remain compatible with the meager (and somewhat controversial) 
observational constraints while leading to very different luminosity functions.

To ascertain the multiplicity statistics of the very low mass stars, we 
are surveying volume-limited samples of nearby
M~dwarfs for companions, combining accurate radial-velocity monitoring with
adaptive optics imaging in the near-infrared.  These
observing techniques together ensure a good sensitivity to stars and brown
dwarfs at all separations (Delfosse et al. \cite{delfosse99b}), as well as
some useful sensitivity to giant planets (Delfosse et al. \cite{delfosse98b}). 
This work was initiated for a 9~pc northern sample (Delfosse et al. 
\cite{delfosse98a}), using the ELODIE spectrograph at Observatoire de 
Haute-Provence (Baranne et al. 1996) and the PUEO adaptive optics system 
at CFHT (Rigaut et al. \cite{rigaut98}). When the HIPPARCOS parallaxes 
became available in 1997 we altered the distance limit to 9.25~pc,
to retain two stars that were already well observed at that point 
(GJ~2066 and Gl~424, none of which is a binary). This also adds two other
stars,  LHS~2784 and the newly identified binary LHS~224.
To further improve the statistics, we now monitor a larger 12~pc 
sample with the FEROS (Kaufer et al.  \cite{kaufer00}), ELODIE and
HARPS (Mayor et al. \cite{mayor03})
spectrographs, and observe it with the NAOS adaptive optics system 
(Rousset et al.  \cite{rousset00}) on the VLT, as well as with PUEO on CFHT. 

Besides these volume limited samples, we observe with the same instruments
two smaller and more loosely defined collections of ``interesting'' 
spectroscopic or astrometric binaries, selected on their apparent 
potential for accurate mass measurements, or in a few cases for having a
possibly substellar companion. The first of these samples was gathered
during a number of published and unpublished CORAVEL programs, while the 
second was culled from the literature. Neither is statistically well defined.

As a result of these programs, we recently published 16~accurate masses 
(S\'egransan et al. \cite{segransan00}), and re-discussed the
Mass-Luminosity relation of the M dwarfs (Delfosse et al. \cite{delfosse00}).  
Here we present stellar components that were newly identified, or newly
resolved, during the adaptive optics observations. Sect. 2 presents the
observing and data-reduction techniques, while section 3 discusses the
properties of the individual new detections.  A statistical analysis of the
northern 9.25~pc sample will be published in a forthcoming paper. 

\begin{figure*}
\begin{tabular}{ccc}
\psfig{height=5.5cm,file=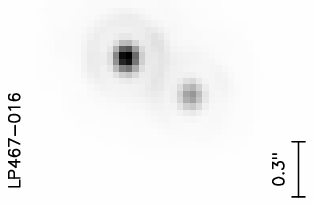,angle=-90} &
\psfig{height=5.5cm,file=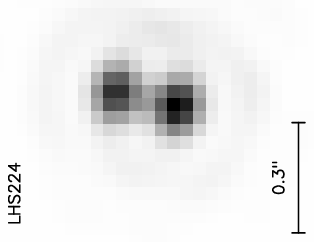,angle=-90} & 
\psfig{height=5.5cm,file=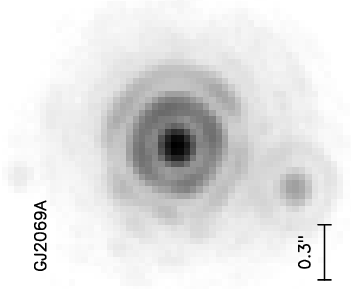,angle=-90} \\
\psfig{height=5.5cm,file=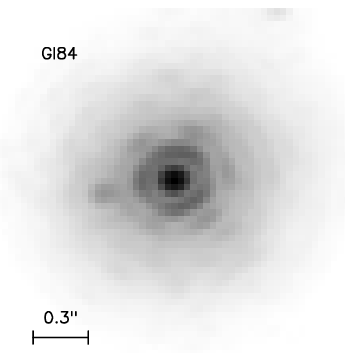,angle=-90} &
\psfig{height=5.5cm,file=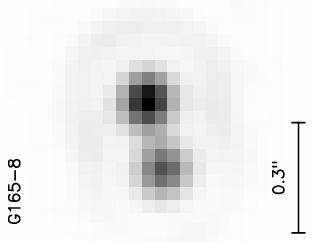,angle=-90} &
\psfig{height=5.5cm,file=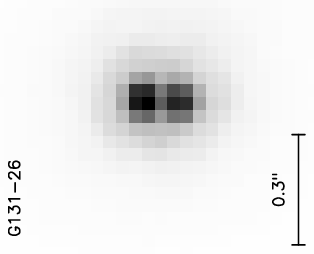,angle=-90} \\
\psfig{height=5.5cm,file=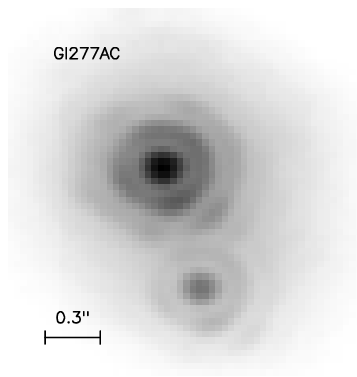,angle=-90} & 
\psfig{height=5.5cm,file=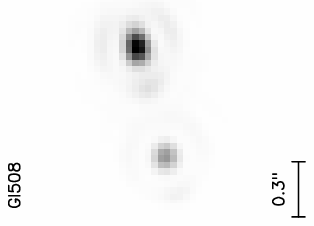,angle=-90} &
\psfig{height=5.5cm,file=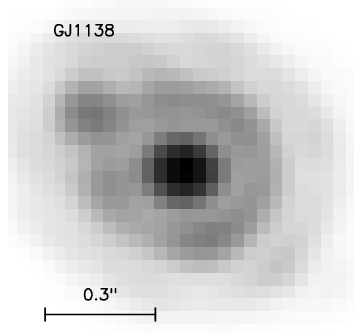,angle=-90} \\
\psfig{height=5.5cm,file=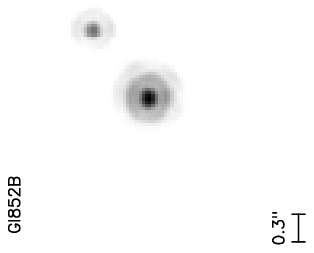,angle=-90} &
\psfig{height=5.5cm,file=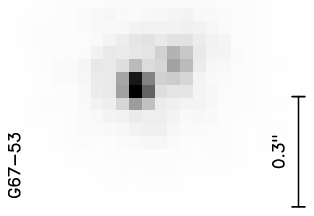,angle=-90} &
\psfig{height=5.5cm,file=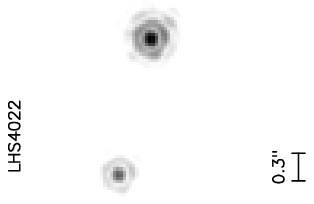,angle=-90} \\ 
\end{tabular}
\caption{Adaptive optics images of new binaries in the volume
 limited samples. The scale
 of each box is indicated by a 0.3'' bar at the bottom left corner. North 
is up and East is left. \label{fig_oa}}
\end{figure*}

\begin{figure*}
\begin{tabular}{ccc}
\psfig{height=5.5cm,file=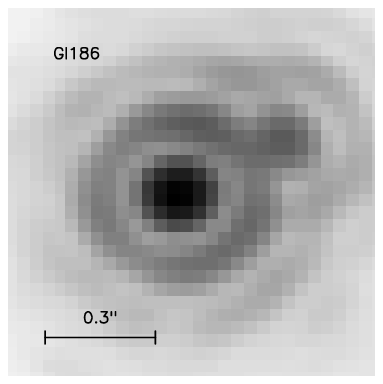,angle=-90} &
\psfig{height=5.5cm,file=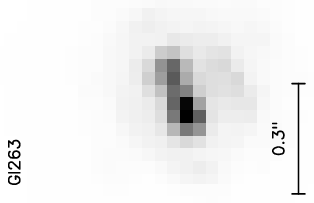,angle=-90} &
\psfig{height=5.5cm,file=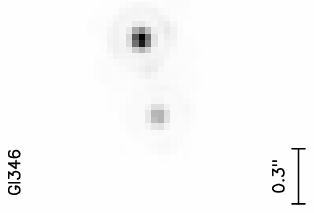,angle=-90} \\
\psfig{height=5.5cm,file=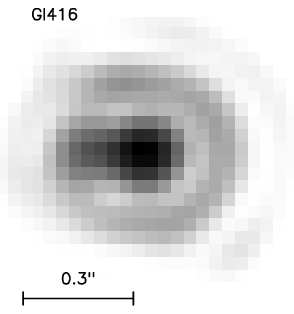,angle=-90} & 
\psfig{height=5.5cm,file=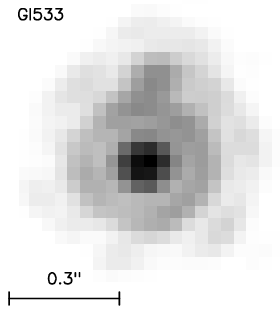,angle=-90} &
\psfig{height=5.5cm,file=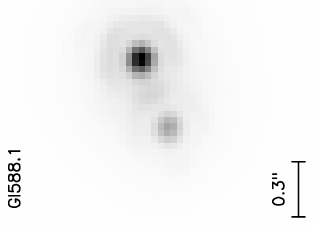,angle=-90} \\
\psfig{height=5.5cm,file=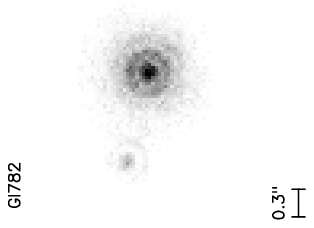,angle=-90} &
\psfig{height=5.5cm,file=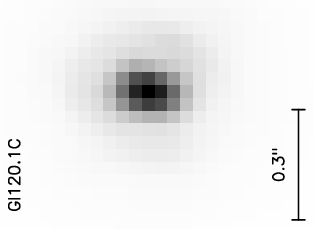,angle=-90} & 
\psfig{height=5.5cm,file=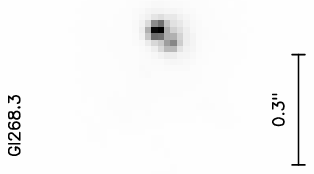,angle=-90} \\
\psfig{height=5.5cm,file=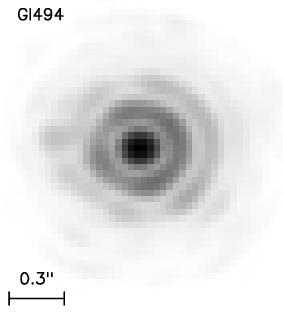,angle=-90}  &
\psfig{height=5.5cm,file=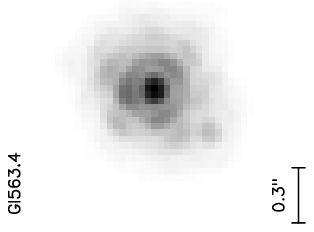,angle=-90} &
\psfig{height=5.5cm,file=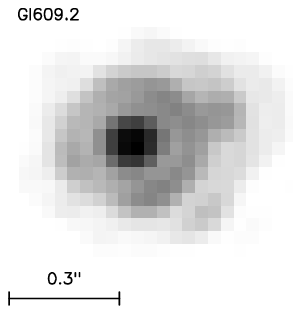,angle=-90} \\ 
 \\
\end{tabular}
\caption{Adaptive optics images of additional newly resolved binaries.
 Scale and orientation are as in Fig.~\ref{fig_oa}. \label{fig_oa_bis}
}
\end{figure*}

\begin{figure*}
\begin{tabular}{ccc}
\psfig{height=5.5cm,file=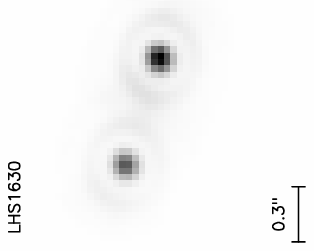,angle=-90} &
\psfig{height=5.5cm,file=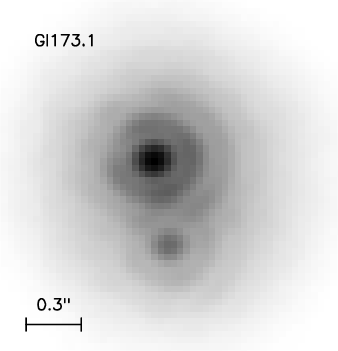,angle=-90} &
\psfig{height=5.5cm,file=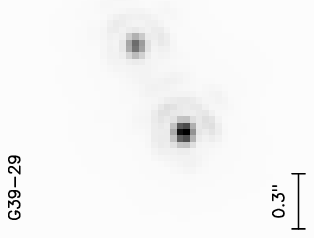,angle=-90} \\
 \\
\end{tabular}
\caption{Continuation of Fig.~\ref{fig_oa_bis}. \label{fig_oa_ter}
}
\end{figure*}

\section{Observations, data reduction and analysis}

\subsection{Instrumental setup}
The observations were carried out at the 3.6-meter Canada-France-Hawaii
Telescope (CFHT) during many observing runs since September 1996,
using the CFHT Adaptive Optics Bonnette (AOB) and two
different infrared cameras. The AOB, also called PUEO after the sharp
vision Hawaiian owl, is a general purpose adaptive optics (AO) system based
on F. Roddier's curvature concept (Roddier et al. \cite{roddier91}). It is
mounted at the telescope F/8 Cassegrain focus, and cameras or other
instruments are then attached to it (Arsenault et al. \cite{arsenault94};
Rigaut et al. \cite{rigaut98}). The atmospheric turbulence is analysed by a
19-element wavefront curvature sensor and the correction applied by a
19-electrode bimorph mirror. The typical control loop bandwidth is 90 Hz at
0 dB. Modal control and continuous mode gain optimization (Gendron \&
L\'ena \cite{gendron94}; Rigaut et al. \cite{rigaut94}) maximize the
quality of the AO correction for the current atmospheric turbulence and
guide star magnitude. For our observations a dichroic mirror diverted the
visible light to the wavefront sensor while a science camera recorded
near-infrared light. We used either MONICA, the Universit\'e de Montr\'eal
Infrared Camera (Nadeau et al. \cite{nadeau94}), or KIR, the CFHT infrared
camera developed to take full advantage of the AO corrected images produced
by PUEO (Doyon et al. \cite{doyon98}).

MONICA was used for the commissioning of the AOB during the first semester
of 1996 and for all science runs until November 1997. It was a facility
instrument based on a NICMOS-3 256 $\times$ 256 detector, and was
originally designed by the Universit\'e de Montr\'eal for the Observatoire
du Mont M\'egantic and CFHT F/8 Cassegrain focii. The camera was retrofitted
with new optics for use at the F/20 output focus of AOB. It produced a
plate scale of 0\farcs034 per pixel, properly sampling diffraction-limited
CFHT images down to the J band (1.25~$\mu$m). The resulting field size is
$8.7\arcsec \times 8.7\arcsec$.

Since December 1997, MONICA has been replaced on PUEO by KIR, an
imaging camera which records a 16 times larger field on an HAWAII 
1024~$\times$~1024 HgCdTe array. KIR also has improved optical quality 
and detector performances, and therefore a better detectivity. The KIR
plate scale is 0\farcs035 per pixel, for a total field size of $36\arcsec
\times 36\arcsec$.

We also observed Gl~268.3 with the Keck~II AO facility in February
2000. The Keck AO system uses a Shack-Hartmann wavefront sensor to measure
the atmospheric distortions and a 349-actuator piezo-stack deformable
mirror to correct for these distortions (Wizinowich et al. 
\cite{wizinowich00a}; Wizinowich et al. \cite{wizinowich00b}). The images
were recorded with the KCAM infrared camera, which has a plate scale of
0\farcs0175 per pixel and an L-shaped field-of-view of 4\farcs5 on a 
long side.

\subsection{Observations}
Sources were first examined for binarity with one filter, usually H (1.65
$\mu$m). Under good to moderate seeing conditions H represents the best
compromise between sensitivity, corrected image quality and sky
brightness. Under worse seeing conditions the Ks filter was used instead to
maintain acceptable image quality. Sources which saturate the detectors in
the minimum available integration time through the H (1.65 $\mu$m) or Ks
(2.23 $\mu$m) broad-band filter (brighter than K~=~7 under typical
conditions) were observed through corresponding narrow-band filters,
usually [Fe$^+$] (1.644 $\mu$m) and either H$_2$ (2.122 $\mu$m) or
Brackett$\gamma$ (2.166 $\mu$m). Whenever a target appeared double or
elongated, it was usually observed with additional filters to
determine relative colour indices. Integration times per frame typically
range between a few tenths of a second and 20 seconds. In order to
improve the signal-to-noise ratio and to average the residual uncorrected
atmospheric turbulence, series of $\sim$4~minute total integration times
were accumulated in a four or five positions mosaic pattern. This observing
sequence also provides for a well determined sky background and a good
correction for detector cosmetic defects. Wavefront sensing was performed
on the sources themselves, which are almost always bright enough (R~$<$~14) to
ensure diffraction-limited images in H and K bands under standard Mauna Kea
atmospheric conditions (i.e.  seeing up to 1\arcsec). The atmospheric
turbulence and AO correction for a given set of observations were
characterized by simultaneously recording the wavefront sensor measurements
and deformable mirror commands, from which an accurate synthetic PSF 
was later generated (Section \ref{data}). Astrometric
calibration fields, such as the central region of the Trapezium Cluster in
the Orion Nebula (McCaughrean \& Stauffer \cite{mccaughrean94}), were
observed to accurately determine the actual detector plate scale and
position angle (P.A.) origin. Flat-field frames were obtained on the 
illuminated dome for each filter. 

The Keck II observations of Gl~268.3 were obtained in H band using a
neutral density filter (ND2, attenuation of $\sim$ 100). A total of 20
individual exposures were co-added to obtain the final 40 s image.
Sky subtraction was performed using images of a close-by field observed
immediately after the science images. No flat-field correction was applied 
for these data.

\subsection{Data reduction and analysis}
\label{data}
For each filter, the raw images were median combined to produce sky frames,
which were then subtracted from the raw data. Subsequent reduction steps
included flat-fielding, flagging of the bad pixels, correction for
systematic detector effects such as the suppression of the remaining $60$
$Hz$ correlated noise, and finally shift-and-add combinations of the
corrected frames into one final image to increase the signal-to-noise
ratio.

For resolved binary systems, the separation, position angle and magnitude
difference between the two stars were then determined using deconvolution
within the AOPHOT software developed by J.-P. V\'eran (\cite{veran98a}). 
For the PUEO observations the long-exposure PSF associated with each 
AO corrected image was reconstructed from the wavefront sensor data and 
deformable mirror commands recorded during each data acquisition. 
The available tools (V\'eran et al. \cite{veran97}; Thomas et al. 
\cite{thomas98}) provide an accurate estimate of the AO-corrected,
long-exposure PSF, when the guide source is of magnitude 13 or brighter. 
No wavefront sensor and deformable mirror data were available for
the Keck observation, and we therefore instead had to generate a 
theoretical estimate of the PSF. This has no adverse consequences for that
particular observation: for such a well resolved and high contrast 
companion the derived parameters are quite insensitive to the adequacy 
of the estimated PSF.
In a second stage, the reconstructed or estimated PSF is used to deconvolve 
the AO image and obtain the pixel coordinates of the primary and secondary
stars, as well as their magnitude difference. V\'eran et al. 
(\cite{veran98b}) provide a complete description of this method and an
assessment of its accuracy.

Application of the astrometric calibrations then yields the desired
parameters. Images of the newly resolved binaries are presented in
Fig. \ref{fig_oa} and Fig. \ref{fig_oa_bis}. 

\subsection{Radial velocity data}
Some of the objects discussed in this paper are also spectroscopic binaries,
identified during long term surveys conducted with the CORAVEL and ELODIE
spectrometers. The radial velocity measurements and their reduction have
been extensively described in Delfosse et al. (\cite{delfosse99b}). 



\begin{table}
\caption{\label{summarize}
  Systems with new low-mass companions in the solar neighbourhood. 
  Spectral types are from Reid et al. (\cite{reid95}), except for Gl~173.1,
  Gl~416, Gl~120.1C, Gl~563.4~, and Gl~609.2 which were taken from SIMBAD.
  They always refer to the combined light of the system.
  Parallaxes (in milliarcseconds) are taken from (h) the
  HIPPARCOS catalogue (ESA \cite{ESA97}), (y) the Yale parallax catalogue 
  (Van Altena et al. 
  \cite{vanaltena95}), (s) S\"oderhjelm \cite{soderhjelm99},
  or are photometric parallaxes from (g) the CNS3  (Gliese and Jahreiss 
  \cite{gliese91}) or (r) spectrophotometric parallaxes from  Reid et al 
  (\cite{reid95}) that we approximately correct for the previously 
  unrecognized multiplicity. Program codes are S9 and S12 for the 
  9.25 and 12pc volume limited surveys, 
  C for the unpublished CORAVEL spectroscopic
  binaries, and L for the spectroscopic and astrometric
  binaries from the literature. A number of objects belong to multiple 
  samples, {\bf and some
  targets from the nominally volume-limited samples have now been found
  to lie at larger distances}.
 }
\begin{tabular}{|l|l@{}l@{}l|l|r|c|} 
\hline
Name         & \multicolumn{3}{c|}{Parallax} &\multicolumn{1}{c|}{Spectral}
    & m$_V$ & Prog.\\ 
             & \multicolumn{3}{c|}{(mas)}    & \multicolumn{1}{c|}{type} & & \\
\hline \hline
G~131--26AB  & 79&$\pm$&14$^{(r)}$           &\hspace{2mm}M4.5 & 13.54 & S12 \\
LP~467--16AB &100&$\pm$&20$^{(r)}$           &\hspace{2mm}M5   & 11.47 & S9 \\
Gl~84        & 105.94&$\pm$&2.04$^{h}$       &\hspace{2mm}M2.5 & 10.19 & LS12\\
Gl~120.1CD   & 38.87&$\pm$&\ \,1.50$^{(h)}$  &\hspace{2mm}K2   &  7.84 & L \\
LHS~1630     & 75&$\pm$&25$^{r}$             &\hspace{2mm}M3.5 & 12.4  & S12 \\
Gl~173.1AabC & 35.21&$\pm$&\ \,1.81$^{(h)}$  &\hspace{2mm}K3   &  9.20 & C \\
G~39-29      & 84&$\pm$&20$^{r}$             &\hspace{2mm}M4   & 12.53 & S12 \\
Gl~186       & 37.93&$\pm$&5.00$^{h}$        &\hspace{2mm}K5   &  8.28 & C  \\
LHS~224AB    & 108.5&$\pm$&\ \,2.1$^{(y)}$   &\hspace{2mm}M5   & 13.30 & S9 \\
Gl~263AB     & 64.96&$\pm$&\ \,2.74$^{(h)}$  &\hspace{2mm}M3.5 & 10.95 & C \\
Gl~268.3AB   & 81.05&$\pm$&\ \,2.42$^{(h)}$  &\hspace{2mm}M2.5 & 10.85 & L \\
Gl~277AC     & 87.15&$\pm$&\ \,4.85$^{(h)}$  &\hspace{2mm}M2.5 & 10.52 & S12C\\
GJ~2069AD    & 78.05&$\pm$&\ \,5.69$^{(h)}$  &\hspace{2mm}M3.5 & 11.89 & S9 \\
Gl~346AB     & 33 &$\pm$&10.0$^{(p)}$        &\hspace{2mm}K7   &  9.70 & C \\
GJ~1138AB    & 102.9&$\pm$&\ \,3.2$^{(y)}$   &\hspace{2mm}M4.5 & 13.02 & S12 \\
Gl~416AB     & 44.00&$\pm$&\ \,1.93$^{(h)}$  &\hspace{2mm}K4   &  9.05 & C \\
Gl~494AB     & 87.50&$\pm$&\ \,1.51$^{(h)}$  &\hspace{2mm}M0.5 &  9.75 & LS12\\
Gl~508AC     & 95.4&$\pm$&\ \,1.6$^{(s)}$    &\hspace{2mm}M0.5 &  8.54 & S12C\\
G165--8AB    & 95&$\pm$&20$^{(p)}$           &\hspace{2mm}M4   & 11.95 & S9 \\
Gl~533AB     & 47.95&$\pm$&\ \,1.77$^{(h)}$  &\hspace{2mm}K7   &  9.80 & C \\
Gl~563.4AB   & 42.26&$\pm$&\ \,1.04$^{(h)}$  &\hspace{2mm}F5   &  5.15 & L \\
Gl~588.1AB   & 34.28&$\pm$&\ \,1.81$^{(h)}$  &\hspace{2mm}K7   & 11.28 & C \\
Gl~609.2AB   & 45.56&$\pm$&\ \,0.89$^{(h)}$  &\hspace{2mm}G8   &  7.10 & L \\
Gl~782AB     & 63.82&$\pm$&\ \,1.49$^{(h)}$  &\hspace{2mm}K4   &  8.92 & C \\
Gl~852BC     & 99.6&$\pm$&\ \,3.3$^{(y)}$  &\hspace{2mm}M5   &  14.4 & S12 \\
G~67--53AB   & 79  &$\pm$&25$^{(r)}$  &\hspace{2mm}M3.5   & 12.10 & S12 \\
LHS~4022AB   & 91  &$\pm$&27$^{(r)}$  &\hspace{2mm}M4   & 11.50 & S12 \\
\hline

\end{tabular}
\end{table}

\section{New companions}

In this section we discuss the properties of the new companions. 
Table~\ref{summarize} summarizes the information available on the systems,
and Table \ref{tab_oa} presents the new adaptive optics information. 

\begin{table*}
\caption{Adaptive optics measurement of the new low-mass companions.
         Gl~268.3 was observed with the Keck adaptive optics system.
         All other observations were obtained at CFHT with the PUEO
         adaptive optics system. Periods listed within parentheses are
         estimated from the observed separation, the distance to the system
         and its approximate mass. They are uncertain by a factor of 
         approximately 3.
         0.5~dex.
  \label{tab_oa}
  }
\begin{center}
\begin{tabular}{|l|c|c|l|c|c|r|c|} \hline
Name & $\rho$  & $\theta$ & $\Delta$m   & Date & Filt. & Period & 
   Accurate mass\\ 
     &  ''     & $\deg$   &             &      &       & (yr) & prospects\\ 
\hline 
G~131--26AB  & 0.111  & 169.9 & 0.46 & 07 Aug 2001 & H & (4)  & Excellent \\ 
LP~467--16AB & 0.409  & 147.2 & 0.69 & 16 Aug 2000 & K & (20) & Fair \\ 
Gl~84        & 0.39   & 101   & 3.59 & 23 Jul 2002 & H & (15) & Fair \\ 
Gl~120.1CD   & 0.083  & 170.1 & 0.02 & 15 Aug 2000 & H & 1.54 & Mediocre\\ 
LHS~1630     & 0.61   & 72    & 0.34 & 18 Sep 2002 & K & (50) & Poor \\ 
Gl~173.1AabC & 0.474  & 188.0 & 1.91 & 05 Mar 1999 & K & (80) & Poor \\ 
G~39-29      & 0.54   & 299   & 0.42 & 13 Sep 2002 & K & (35) & Mediocre \\ 
Gl~186       & 0.31   & 298   & 2.00 & 19-Sep 2002 & K & (40) & Mediocre \\ 
LHS~224AB    & 0.163  & 344.7 & 0.14 & 18 Apr 2000 & K & 3.19 & Good \\ 
Gl~263AB     & 0.110  & 287.4 & 0.46 & 26 Feb 1999 & H & 3.62 & Excellent\\ 
Gl~268.3AB   & 0.054  & 153.6 & 0.57 & 25 Feb 2000 & H & 0.834& Excellent\\ 
Gl~277AC     & 0.684  & 195.0 & 2.02 & 20 Apr 2000 & K & (35) & Mediocre\\ 
GJ~2069AD    & 0.682  & 158.0 & 3.20 & 18 Feb 2000 & K & (40) & Fair\\ 
GJ~2069BC    & 0.549  & 219.1 & 0.52 & 18 Feb 2000 & K & (30) & Fair\\ 
Gl~346AB     & 0.431  & 101.0 & 0.80 & 04 Apr 1999 & H & (60) & Poor\\ 
GJ~1138AB    & 0.303  &  56.1 & 1.89 & 20 Apr 2000 & K & (10) & Good\\ 
Gl~416AB     & 0.138  &  85.0 & 1.05 & 19 Feb 2000 & K & 7.33 & Good\\ 
Gl~494AB     & 0.475  &  81.5 & 4.41 & 19 Feb 2000 & K & 14.5 & Good\\ 
Gl~508AC     & 0.092  & 281.7 & 0.43 & 20 Apr 2000 & K &  1.22& Good\\ 
Gl~508AB     & 0.579  & 102.0 & 0.29 & 20 Apr 2000 & K & 49   & Fair\\ 
G165--8AB    & 0.174  & 253.3 & 0.16 & 19 Feb 2000 & K &  (7) & Good\\ 
Gl~533AB     & 0.230  & 352.8 & 2.02 & 04 Apr 1999 & H & 7.36 & Fair\\ 
Gl~563.4AB   & 0.383  & 139.0 & 3.40 & 26 Feb 1999 & H & 16.07& Fair\\ 
Gl~588.1AB   & 0.408  & 109.6 & 0.97 & 04 Apr 1999 & K & (60) & Poor\\ 
Gl~609.2AB   & 0.220  & 287.0 & 2.04 & 04 Apr 1999 & H & 12.15& Good\\ 
Gl~782AB     & 0.991  &  74.5 & 3.26 & 03 Jul 2001 & K & (100)& Poor\\ 
Gl~852BC     & 0.978  & 305.8 & 1.18 & 07 Aug 2001 & K &  (70)& Poor \\ 
G~67--53AB   & 0.142  & 209.0 & 1.17 & 05 Aug 2001 & J &   (5)& Excellent\\ 
LHS~4022AB   & 1.527  &  74.3 & 1.43 & 05 Aug 2001 & H & (150)& Poor \\ 
\hline
\end{tabular}
\end{center}
\end{table*}

\subsection{New binaries in the volume-limited samples}
\subsubsection{LP~467--16}
Adaptive optics images in August 2000 show a 0.41'' separation pair with a
K-band magnitude difference of 0.69 (Table \ref{tab_oa}).  The
spectroscopic observations also display a $\sim$1\kms drift of the
photocentric radial velocity, as well as variations in the width of the
correlation profile.  From the measured K-band magnitude difference and the
relative slopes of the V and K Mass-Luminosity relations, we estimate a
V-band magnitude difference of $\Delta$(V)$\sim$1. Accounting for the
secondary star therefore changes the photometric parallax from
118$\pm$21~mas, as listed in the CNS3 catalogue, to $\sim$100~mas.  This
pushes the system out of the 9.25~pc volume, but it probably remains within 
12~pc. 

\subsubsection{LHS~224 (G~193-27, L~1750-5)}
LHS~224 is a previously rather anonymous member (only 10 references 
in SIMBAD) of the 9.25~pc sample, with a joint spectral type of M5V
(Reid et al. \cite{reid95}).
We first noticed it as a double-lined spectroscopic binary, with an 
equivalent width ratio of 0.75 in ELODIE spectra. Shortly thereafter, 
an adaptive optics image resolved the system into a 0.16'' binary
with $\Delta$(K)=0.14. {\bf We have now determined a combined 
visual+spectroscopic
orbit for this system, with a period of 3.25~years. The visual part of that 
orbit remains noisy because several observations were obtained under
mediocre weather conditions, and for a 4-m telescope the separation of 
this pair often demands short-wavelength observations. With a few 
additional good measurements at critical phases LHS~224 should provide 
two accurate masses. }

\subsubsection{GJ~2069}

Initially known as a wide ($\sim12''$) visual binary, the GJ~2069 system 
was found to be quadruple by Delfosse et al. (\cite{delfosse99b}). They 
identified GJ~2069A as an eclipsing double-lined spectroscopic binary,
GJ~2069Aab, and resolved a 0.36'' companion to GJ~2069B, GJ~2069C, with a
K-band magnitude difference of 0.45 (Table~\ref{tab_oa} contains a better
measurement on an additional date). Delfosse et al. (\cite{delfosse99a}) 
and S\'egransan et al. (\cite{segransan00}) determined masses with 0.2\%
precision for the two spectroscopic components of GJ~2069A, and suggest
that the system is metal-rich by $\sim$0.5~dex. We now find that adaptive
optics images of the GJ~2069A spectroscopic pair additionally resolve a
fainter companion, GJ~2069D, with $\Delta$(K)=3.2 and $\rho$=0.68'' in
early 2000. The system is thus actually quintuple. 

The extrapolated A-D luminosity contrast in the V band is $\Delta$(V)=4-5,
and detecting D in integrated visible spectra would therefore require a
much better S/N ratio than what we have up to now secured. Eventually the
gravitational pull of GJ~2069D will also cause a measurable drift in the 
systemic velocity of GJ~2069Aab. We have attempted to fit the velocities of
Aa and Ab for this drift, in addition to the elements of the Aab orbit, but 
find that this does not significantly decrease the ${\chi}^2$ of the 
adjusted model. The indeterminacy of this drift indicates that the period
of the Aab-D system is much longer than the present $\sim$6-year
span of our radial-velocity data, consistently with the period estimated
from the separation. In the long run masses can be determined for all 5
components of the GJ~2069 system, {\bf and the radii for the two components 
of the eclipsing pair have been measured (Ribas \cite{ribas03}). The 
system will eventually provide an extremely tight test of low-mass 
theoretical isochrones. }

\subsubsection{G~165-8 (FIRST~J133146.7+291637, LP~323-158, 
1RXS J133146.9+291631)}
With vsini~=~50\kms, G~165-8 is one of only three known M~dwarf ultra-fast 
rotators (vsini~$>$30\kms) within 10~pc (Delfosse et al. \cite{delfosse98a}).
As a consequence of this fast rotation it is magnetically very active, as
illustrated by its figuring in the catalogues of both the FIRST radio and
the ROSAT X-ray surveys. An adaptive optics image shows that it is a binary
with $\Delta$(K)=0.16 and a separation of 0.17'' in early 2000. From this
we infer a V band magnitude difference of $\Delta$(V)=0.25, and correct the
photometric parallax from the CNS3 value of 126$\pm$22~mas to 95$\pm$20~mas. 
G~165-8 therefore most likely does not truly belong to the 9.25~pc volume,
but probably remains within 12~pc. 

Of the two other ultra-fast rotators, Gl~791.2 also is a binary (Hershey 
\cite{hershey78}; Benedict et al. \cite{benedict00}) with a similar
separation. The third ultra-fast rotator, G~188-38, has to date no 
detected companion. Its fast rotation would prevent the radial-velocity 
detection of any close companion, so that a link between fast rotation 
and the presence of a companion remains a possibility. The apparent 
correlation, however, could well represent no more than a fluke of small 
number statistics.


\subsubsection{G~131--26 (LP~131--26, LTT~10045)}
G~131-26, an M4.5 flare star with X-ray and extreme-UV detections, was observed
in August~2001 and found to have a close companion. We have no radial velocity 
information yet, but judging from the magnetic activity of the system the two 
stars are likely to be fast rotators. This would make them less than ideal for 
radial velocity follow-up, and an accurate mass determination would then 
require obtaining an astrometric orbit.

{\bf
\subsubsection{Gl~84 (LHS~149, HIP~9724)}
Gl~84, an M2.5 dwarf at $\sim$10~pc, was identified as a $\sim$15~year
low amplitude single-lined spectroscopic by Nidever et al. 
\cite{nidever02} who derived preliminary orbital elements. 
Our adaptive optics observations resolve the secondary at a $\sim$4~AU
projected separation, and the large contrast from the primary suggests
that it is a late-M dwarf. An unreduced observation confirms the relatively 
rapid movement and the reasonable prospects for mass measurements.
}

{\bf
\subsubsection{LHS~1630 (LP~833-42)}
This M3.5 dwarfs nominally belongs to the 12~pc sample, but our detection
of a fairly bright secondary pushes its photometric parallax somewhat beyond
the distance limit.
}

{\bf
\subsubsection{G~39-29 (LTT~11472)}
The detection of a fairly bright secondary around this magnetically active
M4 dwarfs moves its photometric parallax right to the 12~pc distance limit 
of our extended volume-limited sample. 
}

\subsubsection{Gl~277 (BD+36 1638, LDS 6206)}

Gl 277A is an M2.5 dwarf at a distance of 11.5~pc and forms a 40"
proper-motion pair with the fainter Gl~277B. It is listed in the CNS3 as a
spectroscopic binary, presumably based on the 45\kms spread in eight
M$^t$~Wilson radial velocities from the mid-40s (Abt \cite{abt70}). 11
CORAVEL measurements evenly spread over 6000~days however do show long-term
radial-velocity variations, but with an amplitude of only 4\kms. With 
hindsight, the earlier report probably reflects underestimated
measurement errors for this faint star, even though Gl~277A truly is a 
low-amplitude single-lined spectroscopic binary.
Adaptive optics images in April 2000 resolve it into a 0.68"
pair with a K band contrast of 2 magnitudes. The likely orbital period
is consistent with the spectroscopic lower bound. 

\subsubsection{Gl~508 (ADS~8862, Hu~644)}
The Gl 508 AB pair is a well known long-period orbital binary (P=49~yr, 
a=1.5"), with an observational history that goes back to the beginning of 
the 20$^{th}$ century. 
Individual masses for the two visual components were first determined by 
Heintz (\cite{heintz69}), and the system actually contributed two data points 
to the classic Henry \& Mc~Carthy (\cite{henry93}) Mass-Luminosity paper.
The CNS3 catalogue however notes the primary as a possible spectroscopic 
binary, the Fourth Catalogue of Interferometric Measurements of Binary 
Stars (Hartkopf et al., on-line version
\footnote{http://www.chara.gsu.edu/CHARA/DoubleStars/Spec{\goodbreak}kle/intro.html}) 
similarly mentions tentative evidence for an additional faint component,
and S\"oderhjelm (\cite{soderhjelm99}) remarks that the primary is
overmassive. Gl~508 is indeed seen as a triple-lined spectroscopic
system in many CORAVEL scans, and a preliminary orbit for the inner
pair has P=450~days. 
Gl~508A is also obviously elongated in adaptive optics images
obtained in April 2000, with an estimated contrast of $\Delta$(K)$\sim$0.4
and a separation of approximately 0.1". It is now clear that Gl~508
must be analysed as a triple system. Given the wealth of existing visual 
data for the long-period orbit, the system offers excellent prospects
of shortly determining all three masses.

\subsubsection{GJ~1138 (LHS~293, G119-36)}
GJ~1138 is a previously rather anonymous M4.5V member of the close solar
neighbourhood, at a distance of $\sim$10~pc. April 2000 adaptive optics images
turn out to resolve it into a 0.30'' pair with a K band contrast of
2~magnitudes. An unreduced measurement in 2001 demonstrates a rapid movement,
and confirms the excellent prospects for shortly obtaining two accurate
masses.

\subsubsection{Gl~852 (LDS~782, LHS~3787/3788, Wolf~1561)}
The Gl~852 system, up to now known as a 7'' M4+M5 pair, actually turns out to 
be at least triple. Adaptive optics observations resolve the B component
into a 1'' pair. With a near-IR contrast of only 1~magnitude, the new component
can most likely be detected in uncorrected CCD images under good seeing. Like 
many of the systems where we detect new components, the system is magnetically
very active. This probably indicates a relatively young age, since all 
identified orbital periods are much too long for tidal synchronisation.

\subsubsection{G~67--53 (G~68--5, LTT~16843)}
G~67--53, an M3.5 UV Cet variable and a flare star, is also magnetically
active. It is resolved into a close pair, with one of the shortest estimated
periods amongst the new detections. Its magnetic activity will most likely
hinder precise radial velocity measurements, but astrometric observations
will be able to produce very precise masses.

\subsubsection{LHS 4022 (G~30--23, LTT~17025)}
At a separation of over 1.5'', LHS~4022 is the widest of the new detections,
and within easy reach of visual and CCD observers. We have not yet established
the common proper motion of the two stars, but the absence of any other object
in the 36'' field of view makes their physical association very likely.

\subsection{New binaries from CORAVEL samples}
\subsubsection{Gl~173.1 (HD 286955, HIP 21710)}
The K3-dwarf Gl~173.1A forms a 34'' common proper motion pair with the
M3-dwarf Gl~173.1B. CORAVEL observations show that A is a single-lined
spectroscopic binary with a 610-day period. 
A March 1999 adaptive optics image also resolves it into a 0.47'' pair with
$\Delta$(K)=1.9. At the 28~pc distance of the system, the semi-major axis
of a one solar mass system with the 610~days spectroscopic period is 0.05'', 
well below the observed separation. 
The spectroscopic (Gl~173.1Ab) and adaptive optics Gl~173.1C) companions
of Gl~173.1Aa are therefore different objects, and the Gl~173.1 system is
at least quadruple. The expected period of the AC pair is of the order
of a century, making it of limited interest for mass determinations.
The spectroscopic pair could perhaps be resolved by adaptive optics systems on
8m-class telescopes, but it may need long-baseline interferometric 
observations if its contrast is large.

{\bf
\subsubsection{Gl~186 (HIP 23516, SAO 170021)}
7 CORAVEL radial velocities show that GL~186 is a low amplitude and
long period spectroscopic binary, but they do not cover a full orbital period.
The period estimated from the separation of the resolved companion, 
$\sim$40~years, is compatible with the lower limit on the spectroscopic
period and the two most likely coincide. 
}

\subsubsection{Gl~263 (LHS~1895, HIC 34104)}
28 CORAVEL radial velocities demonstrate that Gl~263 is an eccentric 
spectroscopic binary, with a well determined period of 1313~days (3.6~years) 
and a periastron on JD~2449171. The rest of its orbital elements on the other
hand remain poorly constrained, because observations are lacking
around the critical periastron phase. Adaptive optics images obtained
in February 1999 resolve it into a 0.11'' pair with a 0.45 magnitude 
contrast in the K band. This M3.5V system already provides two preliminary
masses of 0.45 and 0.40, which will become excellent with better radial 
velocity coverage around the periastron. 

\subsubsection{Gl~346 (BD$-$08~2689)}
21 CORAVEL radial velocities demonstrate that Gl~346 is an eccentric 
long-period spectroscopic binary, but they fall short of covering
one orbital period (P$>$14~yr). An April 1999 adaptive optics image
resolves the K7V system into a 0.43'' pair with a K band contrast
of 0.8 magnitude. From this we estimate a V band contrast of $\sim$1.2,
and as a result correct the 40$\pm$10~mas spectro-photometric parallax 
(Reid et al. \cite{reid95}) to 33$\pm$10~mas, nominally pushing Gl~346
beyond the 25~pc limit of the successive Gliese catalogues. 

\subsubsection{Gl~416 (HD 97233, LHS 2370, HIC 54677)}
37 CORAVEL radial velocity measurements determine an accurate 
2670-day (7.3-yr) period for this new K4V spectroscopic binary, but never
separate the two components. A February 2000 adaptive optics image
resolves the system into a 0.14'' pair with a K band contrast of 
1~magnitude, from which we estimate $\Delta$(V)$\sim$1.5~magnitude.
This is small enough that neglecting the secondary light biases
the single-lined orbital elements, which we therefore refrain
from presenting. Modern spectrographs will easily detect the two
components at the next periastron in early 2005, and it will then produce
two accurate masses. 

\subsubsection{Gl~533 (HIP~67808, LHS~2821, BD+13~2721)}
27 CORAVEL observations determine a fairly good 7.3~yr single-lined orbit 
for this new K7V system, even though the phase coverage around periastron is
still somewhat sparse. An April 1999 H~band adaptive optics image 
resolves the system into a 0.23'' pair with $\Delta$(H)=2.0. The extrapolated
contrast in the V band ($\sim$3~magnitudes) is sufficiently large that 
any systematic error stemming from the neglected secondary light are
well below our current random errors on the orbital elements. Spectral
lines from the secondary can probably be detected at the maximum 
radial-velocity separation ($\sim$15\kms) though, and the system would 
then offer excellent prospects for accurate mass measurements.

\subsubsection{Gl~588.1 (HIP 76202, AC +38 34548)}
CORAVEL observations show that this K7V system is an eccentric single-lined 
spectroscopic binary, but they have yet to cover one orbital period 
(P$>$15~yr). An adaptive
optics image in April 1999 resolves the system into a 0.41'' pair with
$\Delta$(K)=1.0. The expected period is consistent with the 
spectroscopic lower bound, strongly suggesting that the same companion
was detected.

\subsubsection{Gl~782 (HIP 99385, HD 191391, LHS 3526)}
CORAVEL observations show that the radial velocity of this well known K4 dwarf 
drifts by $\sim$3\kms over 15~years. An adaptive optics image in July 2001 
resolves the system into a 1'' pair with $\Delta$(K)=3.2. The current angular 
separation and the parallax indicate a very long period, consistent with the 
spectroscopic lower bound.

\subsection{Newly resolved spectroscopic and astrometric binaries}
\subsubsection{Gl~120.1 (HIP 13769, HD 18445~C)}
The Gl~120.1 system associates Gl~120.1AB, a P$\sim$150~yrs visual pair of
K1/K2 dwarfs, with Gl~120.1C, a K2 dwarf, at a projected distance of
$\sim$30''. Duquennoy \& Mayor (\cite{duquennoy91}) found the C component
to be a spectroscopic binary and derived a 1.5-yr single-lined orbit. The
minimum mass of the secondary for this orbit is
M$_2{\times}\sin{i}$=0.042\Msol, well below the limit for stable hydrogen
nuclear burning. Gl~120.1C therefore contributes to the statistical excess
of such companions that Duquennoy \& Mayor (\cite{duquennoy91}) found over
the numbers expected for distributions of companion masses that include no
brown dwarfs. As such it has been discussed in a number of the 
papers (e.g. Mazeh \cite{mazeh99}; Heacox \cite{heacox99}) which examine
the significance of the apparent dearth of close brown-dwarf companions
around solar type stars (the ``brown dwarf desert''), compared with the
larger number of stellar and planetary companion detections.

Halbwachs et al. (\cite{halbwachs00}) recently obtained an astrometric orbit
from the HIPPARCOS intermediate data, which demonstrates that the pair is 
in fact fairly close to face-on, with $\sin{i}{\sim}$0.25. Using the 
Duquennoy \& Mayor (\cite{duquennoy91}) radial-velocity amplitude, they
derive a mass of 0.18~\Msol\  for the spectroscopic companion. This is well
over the maximum mass of a brown dwarf, but still only one quarter of the
primary mass. From Mass-Luminosity relations (e.g. Delfosse et al.,
\cite{delfosse00}) such a star is over three magnitudes fainter in the
near-IR than a K2 dwarf. It was therefore a surprise when adaptive optics
images resolved the system with approximately the expected separation, but
showed that it has a very low contrast (Fig.~\ref{fig_oa}). Because the
pair is not completely separated in Fig.~\ref{fig_oa}, the (zero) magnitude
difference in Table~\ref{tab_oa} is slightly correlated with the measured
separation. Any contrast above $\sim$0.3~magnitude would however produce images
that are very obviously incompatible with the observation. The Gl~120.1CD
pair therefore consists of a pair of early K dwarfs, rather than of a K
dwarf and a mid/late M dwarf. With this additional information, it is now
clear that Gl~120.1C is not a true single-lined system: the Duquennoy \&
Mayor (\cite{duquennoy91}) orbit actually underestimates the actual
velocity amplitude of the primary, because both components of this
low-inclination system contribute a significant fraction of its integrated
light, and yet they are never perceptibly separated at the R$\sim$28~000
resolution of the CORAVEL instruments. It therefore doesn't truly belong to
a sample of single-lined systems with minimum companion masses in the
brown-dwarf range. Its removal obviously weakens the statistical
significance of an excess of such companions, which to date 
represents most of the possible evidence for a population continuity across the
``brown-dwarf desert''. One should note in addition that Halbwachs et al.
(\cite{halbwachs03})
exclude a continuity between stellar and planetary 
companions to solar-type stars, through an analysis of a sample that 
includes the Duquennoy \& Mayor (\cite{duquennoy91}) objects but
benefits from additional CORAVEL observations and from a larger
and cleaner sample. 
The case
for a population in the ``brown-dwarf desert'' would be further weakened 
if the Duquennoy \& Mayor (\cite{duquennoy91}) sample and its
Halbwachs et al. (\cite{halbwachs03}) update turned out to still contain 
a small number of unrecognized double-lined systems, as could conceivably 
be the case. We are
currently obtaining data with the ELODIE and CORALIE spectrographs to
settle this issue.

\subsubsection{Gl~268.3 (HIP 35191, BD +27 1348)}
Gl~268.3 was found to be a 304-day double-lined spectroscopic binary by 
Delfosse et al. ({\cite{delfosse99b}). They also presented marginally
resolved CFHT adaptive optics images but could not determine quantitative
parameters. The pair is well resolved in a February 2000 Keck adaptive optics
image, with $\Delta$(H)=0.57, rho=0.054'' and theta=123.5$\deg$. Preliminary 
masses of 0.49 and 0.33 solar masses can be derived from the presently 
available 
data, and a few additional adaptive optics observations should pinpoint
them with excellent accuracy. 

\subsubsection{Gl~494 (DT~Vir, LHS~2665, HIP~63510, BD+13~2618)}
Gl~494A is a magnetically very active M0.5 dwarf, as a result of its fast 
rotation (v~sin~{\it{i}}~=9.6\kms, determined with ELODIE). It
is not a tidally-locked short-period binary, and the age--rotation relation
therefore implies that it is fairly young, most likely younger than 1~Gyr.
Heintz (\cite{heintz90}; \cite{heintz94}) obtained a convincing 14.5-yr 
astrometric orbit for Gl~494, which established that the M0.5V primary has a 
much lower-mass companion. 26 CORAVEL radial-velocity measurements 
show $\pm$1.5\kms variations at approximately this 
orbital period, and therefore validate the astrometric detection. Heintz 
asserted the 
companion to be substellar, based however on an estimated mass of 0.4\Msol\ 
for an M2V primary. Gl~494A has since been reclassified as M0.5V (Reid et al.
\cite{reid95}), so that 0.6\Msol\  is now a more appropriate guess for its
mass. This rescaling brings the Heintz companion mass close to the 
substellar limit.

Henry et al. (\cite{henry99}) did not detect the companion in either of
two {\it HST} observations with FGS3 used in TRANS mode . At the $\sim$0.4'' 
separation, expected from the astrometric orbit and an estimated 
0.6\Msol\  mass for the M0.5V primary, this implies a V band contrast of at 
least 3.5~magnitudes and confirms that the companion is indeed rather faint.
K band adaptive optics images in February 2000 resolve the system into a 
0.48'' pair, with $\Delta$(K)=4.4. With M$_K$=9.7, its spectral type is 
approximately M7V (Leggett \cite{leggett92}). The extrapolated contrast in 
the V band is very large, and does explain why the FGS observations could 
not detect the companion. It should on the other hand be within reach of 
{\it HST} imaging observations.

 From the orbital elements of the astrometric orbit of Heinz (\cite{heintz94})
and the observed separation in April 1999, we can in principle determine a 
scale factor between the astrometric and relative orbits of 10.8. From this 
a semi-major axis of 0.55'' immediately follows for the relative orbit. 
Together with the HIPPARCOS parallax this results in a system mass of 
1.19\Msol, which the scale factor then splits into 1.08\Msol\  for the 
M0.5V primary and 0.11\Msol\  
for the faint secondary. The elements of the Heintz (\cite{heintz94}) 
astrometric orbit are quoted without standard errors, so that we are not
in a position to evaluate precise confidence intervals.
Still, 1.08\Msol\  cannot possibly be the mass of a single M0.5 dwarf, and we
also note an 80~degrees discrepancy for the predicted position angle. We 
therefore suspect that either the actual confidence intervals are so wide as 
to be of little use, or perhaps there is a problem in the Heintz elements, for
instance because they force a circular orbit for a system which may have
some small eccentricity.

For the time being we therefore prefer to turn to its absolute magnitude to 
evaluate the position of the secondary relative to the substellar 
limit. From the Baraffe et al. (\cite{baraffe98}) models,
an object of  M$_K$=9.7 has to be younger than 300~Myr to be 
a brown dwarf. If it has instead reached the main sequence, its mass for the
same models would be 0.09\Msol.
The age range that can be inferred from the characteristics of its primary
is at present insufficiently constrained to presently ascertain whether 
Gl~494B is a 
brown dwarf or a star, but whatever its status it will clearly play an
important role in characterizing the substellar transition.
Together with the Gl~569 (Mart\'{\i}n et al. \cite{martin00}) and GJ~2005
(Leinert et al. \cite{leinert94}) systems, Gl~494 contains one of the 
faintest known objects for which dynamical masses can be obtained on realistic
timescales. Continued
observations with adaptive optics, astrometric satellites, and radial-velocity
spectrographs, have together the potential to determine its mass from first 
principles and with excellent accuracy. 

\subsubsection{Gl~563.4 ($\alpha$1 Lib, FK5 1387, HIP 72603, HD 130819)}
Gl563.4, a V=5.1 F3 dwarf, forms a common proper motion system with the 
brighter (V=2.8) A3-dwarf Gl 564.1. The Hipparcos proper motions of the 
two stars are mildly discrepant ((-0.136,-0.059) vs (-0.106,-0.069)), but 
at a level that is consistent with the orbital motion of the Gl~563.4 
photocenter over the Hipparcos mission.
Duquennoy \& Mayor (\cite{duquennoy91}) found it to be
a single-lined spectroscopic binary and derived a preliminary orbit.
With additional CORAVEL measurements obtained since 1991, the 
orbit is now definitive and has P~=~5870~days.
Adaptive
optics images obtained in February 1999 resolve Gl 563.4 into a 
0.38'' pair, with $\Delta$(H)=3.4. It most likely coincides with the 
spectroscopic system and offers good prospects of determining reasonably
accurate masses. 

\subsubsection{Gl~609.2 (HD 144287, HIP 78709, BD+25 3020)}
Duquennoy \& Mayor (\cite{duquennoy91}) found Gl~609.2, a G8 dwarf, to be 
a spectroscopic binary and determined a 12-yr single-lined spectroscopic 
orbit. Multiple speckle observations with large telescopes failed to 
resolve this pair (Bonneau et al. \cite{bonneau86}; Mason et al. 
\cite{mason99}), in retrospect because of its large contrast 
in visible bands. Adaptive optics images obtained in april 1999 easily 
resolve the system into a 0.21'' pair, but with a $\Delta$(H)=2.2 infrared
contrast. This translates into values for visible bands that are 
somewhat beyond
the typical dynamic range of speckle observations. The Gl~609.2 system will 
provide accurate masses if spectral lines from the faint secundary can be
detected, as might be possible with improved analysis tools (e.g.
Zucker et al. \cite{zucker03}).

\begin{acknowledgements}
We thank the technical staff and telescope operators of CFHT for their 
support during these long-term observations. J.-L. B. would also like to 
thank Meghan Gray and Diana Chaytor who reduced part of the AO data while 
being students at CFHT in 1997 and 1998.

"This research has made extensive use of the SIMBAD database,operated at 
CDS, Strasbourg, France, and of the Washington Double Star Catalog maintained 
at the U.S. Naval Observatory."
\end{acknowledgements}

\end{document}